# Speech frame implementation for speech analysis and recognition

A.A. Konev, V.S. Khlebnikov, A.Yu. Yakimuk

Tomsk State University of Control Systems and Radioelectronics, 40 Lenina Ave., Tomsk, 634012 Russia
E-mail: kaa1@keva.tusur.ru

**Abstract.** Distinctive features of the created speech frame are: the ability to take into account the emotional state of the speaker, support for working with diseases of the speech-forming tract of speakers and the presence of manual segmentation of a number of speech signals. In addition, the system is focused on Russian-language speech material, unlike most analogs.

**1. Introduction**

The creation of systems for automatic speech processing is one of the most urgent directions in the development of modern information technologies. This area includes speech recognition in various languages, keyword search, text-dependent and text-independent speaker identification systems, language identification systems, assessment of the quality of speech recordings, voice change systems. Any of the systems listed above requires speech material for analysis, training, or algorithm development.

The aim of this work is to design and implement a speech frame that meets the needs of modern speech analysis and recognition systems.

The speech frame should include speech signals and a database containing their structured description, as well as automatic processing tools that allow the user to search for certain parameters of sound signals.

**2. Speech frame design.**

The infological level of design is an information-logical model of the subject area, from which redundancy of data is excluded and informational features of the control object are displayed without taking into account the features and specifics of a particular DBMS.

Here is a description of of the speech frame and the basic requirements for its design, as the data necessary for building an information-logical model. There are 27 entity tables in total. The tables "Speech signals", "Announcers", "Sound units" and "Segmentation" occupy the central place in the structure of the database, the rest of the tables serve as reference books for various parameters. A model of the physical structure of the database was created, taking into account the rules of cascade update and cascade deletion.

A glossary of entities with a brief description of the stored information is given in the form of Table 1.

Table 1. Glossary of Entities

| Entity name | Description |
|---|---|
| ACOUSTIC_ENVIRONMENT | Acoustic environment information |
| BOOK_DEFECTS | Description of Spelling Defects |
| BOOK_DIALECTS | Description of dialects and languages |
| BOOK_EMOTIONS | Emotion data |
| BOOK_LABIALIZATION | Labialization vowel classifier |
| BOOK_LOCATION | Place of origin classifier |
| BOOK_RISE | Vowel Ascent Classifier |
| BOOK_ROW | Vowel classifier by row |
| BOOK_SEX | Speakers gender description |
| BOOK_SOFT | Hardness consonant classifier |
| BOOK_SPEECH_TEMPS | Speech rate data |
| BOOK_SPEECH_TYPES | Speech type data |
| BOOK_STRESSED | Description of stress types |
| BOOK_UNIT_TYPES | Types of speech units |
| BOOK_VOICED | Description of voicing / hardness |
| BOOK_VOICE_TYPES | Voice type information |
| BOOK_WAY_OF_ORIGIN | Classifier by the way of consonant formation |
| CLASS | Description of sound units |
| COMMUNICATION_CHANEL | Description of the communication channel |
| FILE_FORMAT | File format description |
| NOISE | Synthetic Noise Information |
| RECORDING_DEVICE | Recorder data |
| SEGMENTATION | Segmentation |

A glossary of attributes for each entity with a description of primary keys (PK - Primary Key) and foreign keys (FK - Foreign Key) is given in tables 2-18. The ACOUSTIC_ENVIRONMENT table contains information about possible types of acoustic environment at the time of recording a speech signal. The acoustic environment can be: office space (maximum noise level 20 dB), car interior (noise level 40 dB), etc.

Table 2. Glossary of attributes for the ACOUSTIC_ENVIRONMENT table

| Attribute name | Description | Data type |
| --- | --- | --- |
| ENVIRONMENT_ID(PK) | Environment code | Integer |
| NOISE_LEVEL(DB) | Noise level in decibels | Floating point |
| TITLE | Acoustic environment name | Text |

The BOOK_DEFECTS table contains information about speech defects encountered by speakers. It is a reference table, the structure of the database allows you to add new attributes to this table for a more detailed description of defects.

Table 3. Glossary of attributes for the BOOK_DEFECTS table

| Attribute name | Description | Data type |
| --- | --- | --- |
| ID_DEFECT (PK) | Defect code | Integer |
| TITLE | Defect name | Text |

The BOOK_DIALECTS table contains information about the dialects of different languages. For the Russian language, guided by research within the framework of the SpeechDat project, it was decided to use the following distribution of speakers by regional characteristics (dialects):

– Moscow and St. Petersburg;
– South of Russia;
– North of Russia;
– the Urals, Siberia and the Far East;
– The central part of Russia.

Table 4. Glossary of attributes for the BOOK_DIALECTS table

| Attribute name | Description | Data type |
| --- | --- | --- |
| ID_DIALECT (PK) | Dialect code | Integer |
| TITLE | Dialect name | Text |
| LANGUAGE | The name of the language to which this dialect belongs | Text |

The BOOK_EMOTIONS table contains possible variants of the emotional coloring of speech signals. After studying a number of works in this area, it was decided to use the following classification of the speaker's emotional state, based on the work [1] (Table 5).

Table 5. Glossary of attributes for the BOOK_EMOTIONS table

| Attribute name | Description | Data type |
| --- | --- | --- |
| ID_EMOTION (PK) | Emotion code | Integer |
| TITLE | Name | Text |

The BOOK_LABIALIZATION table contains the classification of vowels by labialization ("rounding" when pronouncing). The vowels O and U are classified as labialized phonemes, the rest of the vowels are considered non-labialized. The BOOK_LOCATION table stores a description of the classification of consonants according to the place of formation. In the developed speech frame, the classification of consonants according to voicedness / voicelessness, softness / hardness, place of formation and method of formation in accordance with the work is implemented.

Table 6. Glossary of attributes for the BOOK_ LABIALIZATION table

| Attribute name | Description | Data type |
| --- | --- | --- |
| ID (PK) | Class code | Integer |
| TITLE | Name | Text |

Table 7. Glossary of attributes for the BOOK_LOCATION table

| Attribute name | Description | Data type |
| --- | --- | --- |
| ID (PK) | Location code | Integer |
| TITLE | Name | Text |

The BOOK_ RISE table contains the possible types of lifting of the tongue when pronouncing vowels (upper, middle and lower ascents).

With the help of the BOOK_ ROW table, the division of vowels into types by rows is realized.

Table 8. Glossary of attributes for the BOOK_RISE table

| Attribute name | Description | Data type |
| --- | --- | --- |
| ID (PK) | Lift type code | Integer |
| TITLE | Name | Text |

Table 9. Glossary of attributes for the BOOK_ ROW table

| Attribute name | Description | Data type |
| --- | --- | --- |
| ID (PK) | Row code | Integer |
| TITLE | Name | Text |

The BOOK_SEX table contains information about the speaker field. The speech frame does not provide for support for gender-reassigned speakers, due to the specific features of the voice of such speakers.

The BOOK_SOFT table contains a list of possible values for the softness of a sound unit (hard consonant, soft consonant, sonorant).

Table 10. Glossary of attributes for the BOOK_SEX table

| Attribute name | Description | Data type |
| --- | --- | --- |
| ID (PK) | Gender code | Integer |
| TITLE | Name | Text |

Table 11. Glossary of attributes for the BOOK_SOFT table

| Attribute name | Description | Data type |
| --- | --- | --- |
| SOFT_ID (PK) | Code | Integer |
| TITLE | Name | Text |

The BOOK_SPEECH_TEMPS table contains information about possible speech rates. The following classification is provided:
- normal pace (up to 8 sounds per second inclusive);
- accelerated tempo (up to 12 sounds per second inclusive);
- fast pace (more than 12 sounds per second. Average value of 20 sounds) [2].

Table 12. Glossary of attributes for the BOOK_SPEECH_TEMPS table

| Attribute name | Description | Data type |
| --- | --- | --- |
| ID (PK) | Speech rate code | Integer |
| SPEED | Name | Text |
| SOUNDS_PER_ SECOND | Speed (number of sounds per second) | Integer |

The BOOK_STRESSED table contains information about the possible types of stress.

In the created alphabet of sound units, all sound units are divided into classes: vowels and consonants. For consonants, the table shows the absence of stress. Vowels, in turn, are divided into subclasses: percussive and unstressed. It should be noted that gradation exists within these subclasses as well. A significant role in the sound of a stressed vowel is played by the sound units surrounding it. In this regard, this subclass is divided into four groups:
1 - between solid;
2 - between hard and soft;
3 - between soft and hard;
4 - between soft.

An important factor for unstressed vowels is the type of consonant in front of it and the position relative to the stressed vowel. In this regard, the Potebnya formula was used: the vowels of the stressed syllable - 3 units, the first pre-stressed - 2, the second pre-stressed and post-stressed - 1 unit. According to this formula, unstressed vowels are divided into the following groups: strength of 2 units after hard, strength of 1 unit after hard, strength of 2 units after soft, strength of 1 unit after soft.

Thus, there are 4 options for stressed vowels, 5 options for unstressed ones, 1 option for consonants and 1 option-exception for marking the pause between sounds.

Table 13. Glossary of attributes for the BOOK_SPEECH_TEMPS table

| Attribute name | Description | Data type |
| --- | --- | --- |
| ID_STRESSED (PK) | Accent type code | Integer |
| TITLE | Description of stress | Text |

The BOOK_UNIT_TYPES table contains information about the types of speech units. The following types of speech are considered - syllable, phrase, text, sound.

The BOOK_VOICED table contains a list of possible voicing values of a sound unit (voiceless consonant, voiced consonant, vowel).

Table 14. Glossary of attributes for the BOOK_UNIT_TYPES table

| Attribute name | Description | Data type |
| --- | --- | --- |
| TYPE_ID (PK) | Speech unit type code | Integer |
| TITLE | Name | Text |

Table 15. Glossary of attributes for the BOOK_VOICED table

| Attribute name | Description | Data type |
| --- | --- | --- |
| VOICED_ID(PK) | Code | Integer |
| TITLE | Name | Text |

The BOOK_VOICE_TYPES table contains information about the types of voice. The following types of voice are considered - talking, singing, whispering, esophageal.

The BOOK_WAY_OF_ORIGIN table contains the classification of consonants by origin. The consonants are divided into occlusive plosive, occlusive affricates, occlusive nasal and slotted types.

Table 16. Glossary of attributes for the BOOK_VOICE_TYPES table

| Attribute name | Description | Data type |
| --- | --- | --- |
| ID (PK) | Voice type code | Integer |
| TITLE | Name | Text |

Table 17. Glossary of attributes for the BOOK_WAY_OF_ORIGIN table

| Attribute name | Description | Data type |
| --- | --- | --- |
| ID (PK) | Voice source type code | Integer |
| TITLE | Name | Text |

The CLASS table contains information about sounds. The classification is introduced in accordance with the developed alphabet for the sound units of the Russian language. A total of 77 sound units in the Russian version of the sound alphabet. For a foreign language, a new alphabet must be introduced. The WAY_OF_ORIGIN, LOCATION fields are filled only for consonants, the LABIALIZATION, RISE, ROW fields are only for vowels [3].

Table 18. Glossary of attributes for the BOOK_SPEECH_TEMPS table

| Attribute name | Description | Data type |
| --- | --- | --- |
| SYMBOL (PK) | Symbolic designation of a sound unit | Integer |
| STRESSED (FK) | Accent mark | Text |
| VOCALIZED | Sign of vocalization | Integer |
| SOFT(FK) | Sign of softness | Integer |
| VOICED(FK) | Sign of voicing | Integer |
| LOCATION (FK) | Place of education | Integer |
| WAY_OF_ORIGIN (FK) | Method of education | Integer |
| LABIALIZATION (FK) | Labilization | Integer |
| RISE (FK) | Raising the tongue when making a sound | Integer |
| ROW (FK) | Vowel row | Integer |

The COMMUNICATION_CHANNEL table contains data about the used communication channels. In the general case, the use of a communication channel is not mandatory, but in the case of use in the corresponding field of the speech signal description, a link is given to one of the entries in the COMMUNICATION_CHANNEL table.

The RECORDING_DEVICE table contains the description and characteristics of the voice recording device (microphone, mobile phone, and so on).

Table 19. Glossary of attributes for the COMMUNICATION_CHANNEL table

| Attribute name | Description | Data type |
| --- | --- | --- |
| ID (PK) | Channel code | Integer |
| TITLE | Name | Text |

Table 20. Glossary of attributes for the RECORDING_DEVICE table

| Attribute name | Description | Data type |
| --- | --- | --- |
| DEVICE_ID (PK) | Recorder ID | Integer |
| TYPE | Device type | Text |
| BANDWIDTH | Bandwidth | Float |

The NOISE table contains a description of synthetic noise. By default, in the SPEECH_SIGNAL table, the NOISE_ID attribute is set to 0 (no noise), if desired (the presence of synthetic noise in the speech signal), you can set a link to one of the entries in the NOISE table.

The FILE_FORMAT table contains a description of the characteristics of the file formats used for recording speech signals. The table gives a description of the used bit rate, sampling rate, number of channels and file type [4].

Table 21. Glossary of attributes for the FILE_FORMAT table

| Attribute name | Description | Data type |
| --- | --- | --- |
| ID (PK) | File type code | Integer |
| DISCRETIZATION_FREQUENCY | Sampling frequency | Float |
| BITRATE | Bitness | Integer |
| FILE_TYPE | File type | Text |
| NUMBER_OF_CHANNELS | Number of channels | Integer |

Table 22. Glossary of attributes for the NOISE table

| Attribute name | Description | Data type |
| --- | --- | --- |
| ID_NOISE (PK) | Synthetic noise code | Integer |
| NOISE_TYPE | Name or description of the noise | Text |
| SIGNAL/NOISE_RATIO(DB) | Signal to noise ratio in decibels | Float |

The SEGMENTATION table contains the results of segmentation (division of continuous speech into a sequence of sound units). The table can store both the results of manual segmentation and automatic. To control the quality of automatic segmentation, the speech frame contains the results of manual segmentation of seven speech units (103 speech signals in total). Segmentation of each phrase was checked by at least two experts. The signals are selected in such a way that each character of the sound alphabet occurs in manual segmentation at least three times [5].

Table 23. Glossary of attributes for the SEGMENTATION table

| Attribute name | Description | Data type |
| --- | --- | --- |
| POSITION(PK) | The position of the sound unit in the entire speech signal | Integer |
| FILENAME (PK)(FK) | The name of the speech signal containing this segment of segmentation | Integer |
| START_AUDIO | Beginning of a segment of the segmentation | Float |
| TYPE_ID(FK) | Sound unit code | Text |

The SICKNESS table contains information about diseases of the vocal tract. The structure of the table allows you to add new attributes describing speech disease.

The SPEAKER table contains information about the speakers. The records of the dates of birth (to calculate the age of the speaker), dialectical affiliation, gender of the speakers are kept.

The SPEECH_SIGNAL table contains a detailed structured description of each signal contained in the speech frame. A description is given of the file structure of the speech signal, the acoustic environment during recording, the characteristics of the recording device, the communication channel, synthetic noises (if any), the voice type of the speaker, the type of speech and voice used, the rate of speech, diseases of the vocal tract, accent and speech defects.

The SPEECH_UNIT table contains information about speech units. The table contains the spelling of a speech unit, information about its type, as well as transcription in the selected sound alphabet.

Table 24. Glossary of attributes for the SICKNESS table

| Attribute name | Description | Data type |
|---|---|---|
| ID_SICKNESS(PK) | Disease code | Integer |
| TITLE | Name | Text |

Table 25. Glossary of attributes for the SPEAKER table

| Attribute name | Description | Data type |
|---|---|---|
| ID(PK) | Announcer code | Integer |
| SEX(FK) | Floor | Integer |
| NAME | Name | Text |
| SURNAME | Surname | Text |
| FAMILY_NAME | Surname | Text |
| BIRTH_DATE | Date of Birth | Date |

Table 26. Glossary of attributes for the SPEECH_SIGNAL table

| Attribute name | Description | Data type |
|---|---|---|
| FILE_NAME (PK) | File name | Text |
| SPEECH_UNIT_ID (FK) | Sound unit code | Integer |
| FILE_NAME | File name | Text |
| LENGTH | Speech signal length | Time |
| RECORD_DATE | Date of recording | Date |
| FILE_FORMAT (FK) | File format description code | Integer |
| SYNTHETIC_NOISE_ TYPE (FK) | Synthetic noise code | Integer |
| RECORDING_DEVICE (FK) | Recorder code | Integer |
| DIALECT_ID (FK) | Dialect code | Integer |
| ACOUSTIC_ ENVIRONMENT(FK) | Acoustic environment code | Integer |
| SPEECH_TYPE_ID (FK) | Speech type code | Integer |
| VOICE_TYPE_ID (FK) | Voice type code | Integer |
| SPEECH_TEMP_ID (FK) | Speech rate code | Integer |
| CHANNEL (FK) | Transmission channel code | Integer |
| SPEECH_SICKNESS (FK) | Speech disease code | Integer |
| ACIENT | The presence of an accent | Integer |
| SPEECH_DEFECT (FK) | Speech Defect Code | Integer |
| EMOTIONAL_STATE (FK) | Emotional state code | Integer |
| SPEAKER_ID (FK) | Announcer code | Integer |

Table 27. Glossary of attributes for the SPEECH_UNIT table

| Attribute name | Description | Data type |
|---|---|---|
| ID(PK) | Announcer code | Integer |
| SPELLING_RECORD(FK) | Spelling notation | Text |
| TRANSCRIPTION | Transcription | Text |
| UNIT_TYPE | Speech unit type | Integer |

Within the framework of this section of the work, the basic requirements for the structure of the speech frame and the structure of tables were formulated. All requirements were reflected in the conceptual database model. Based on this model and the described business rules, the physical implementation of the database was carried out.

**3. Preparation of speech material.**

The following requirements were imposed on the selection of speakers:

- the presence of male and female speakers;
- the presence of people of different age categories;
- the presence of speakers with speech impairments.

In accordance with the given requirements, 193 speakers were selected (49 male speakers and 144 female speakers).

Based on a conceptual data model and an attribute level diagram, a speech database has been implemented. The main table of the speech frame is the SPEECH_SIGNAL table, this table contains the description of speech signals. The rest of the tables contain the data necessary to describe one or another parameter of the speech signal. The exception is tables SPEAKER, CLASS and SEGMENTATION, which contain data on speakers, sound units and manual segmentation, respectively.

As a result of the work carried out, the speech frame contains:
- 77 speech units;
- data on 193 speakers;
- 124 speech signals with a total duration of 842 seconds (14 minutes 2 seconds);
- results of manual segmentation for 103 speech signals.

One of the requirements in the design of the speech frame was the ability to filter records with the description of speech signals according to the characteristics specified by the user. This requirement was implemented using queries built in SQL (Structured Query Language).

The user of the speech frame is given the opportunity to select several parameters for the search. Processing requests of this type is implemented as follows:
1. Using the first search parameter, the speech frame processes the request and generates a sample of signals from the database.
2. using the second search parameter, a new request is processed, but the signals are sampled not from the database, but from the sample of signals generated at the previous stage.
3. Stage 2 is repeated until the search parameters are over.
4. the results of the work are displayed to the user on the screen.

In addition to requests for data samples from the speech frame, requests for calculating statistical information about the speech frame were implemented.

The correctness of all SQL queries was tested using the black box method.

## 4. Conclusion

Within the framework of this work, a speech frame was created, which includes speech signals and a database containing their structured description. Distinctive features of the created speech frame are: the ability to take into account the emotional state of the speaker, support for working with diseases of the speech-forming tract of speakers and the presence of manual segmentation of a number of speech signals. In addition, the system is focused on Russian-language speech material, unlike most analogs.

At the moment, the speech frame contains: 77 speech units, data on 193 speakers, 124 speech signals with a total duration of 842 seconds, the results of manual segmentation for 103 speech signals. The speech database consists of 27 tables; 37 SQL queries have been implemented to implement the search functionality in the database.

## 5. Acknowledgments

This research was funded by the Ministry of Science and Higher Education of Russia, Government Order for 2020–2022, project no. FEWM-2020-0037 (TUSUR).

**References**
[1] Devillers L., Vidrascu L. 2007 *Speaker Classification II* **4441** 34-42
[2] Kostuchenko E, Novokhrestova D, Pekarskikh S, Shelupanov A et al. 2019 *SPECOM 2019: Speech and Computer* **11658** pp 359-369
[3] Reddy A, Subramanian 2015 *Journal of Voice* **29(5)** pp 603–610
[4] Deutsch D and Hamaoui K 2005 *Acoustical Society of America* **117** 2476
[5] Yakimuk A Yu, Konev A A, Andreeva Yu V, Nemirovich-Danchenko M M 2019 *IOP Conference Series: Materials Science and Engineering* **597(1)** 012072.